\documentclass[conference]{IEEEtran}
\IEEEoverridecommandlockouts

\usepackage{cite}
\usepackage{amsmath,amssymb,amsfonts}
\usepackage{algorithmic}
\usepackage{graphicx}
\graphicspath{{./}{../}}
\usepackage{textcomp}
\usepackage{xcolor}
\usepackage{multirow}
\usepackage{url}
\def\BibTeX{{\rm B\kern-.05em{\sc i\kern-.025em b}\kern-.08em
    T\kern-.1667em\lower.7ex\hbox{E}\kern-.125emX}}
\begin{document}
\pagestyle{plain}

\title{LWM-Spectro: A Foundation Model for Wireless Baseband Signal Spectrograms\\
}

\author{Namhyun Kim, Sadjad Alikhani, and Ahmed Alkhateeb\\
School of Electrical, Computer, and Energy Engineering, Arizona State University\\
Email: {namhyun, alikhani, alkhateeb}@asu.edu}

\maketitle

\begin{abstract}
The received in-phase and quadrature (I/Q) baseband signals inherently encode physical-layer and channel characteristics of wireless links.
Learning robust and transferable representations directly from such raw signals, however, remains challenging due to heterogeneous communication systems, diverse propagation environments, and limited labeled data.
To address this, we present \textit{LWM-Spectro}\footnote{\url{https://huggingface.co/wi-lab/lwm-spectro}}, a transformer-based foundation model pretrained on large-scale I/Q data represented as time--frequency spectrograms.
The model leverages self-supervised masked modeling, contrastive learning, and a mixture-of-experts (MoE) architecture to learn general-purpose wireless representations.
These representations transfer effectively to downstream tasks such as modulation classification and joint SNR/mobility recognition, even with minimal supervision.
Across tasks, LWM-Spectro consistently outperforms state-of-the-art deep learning baselines in both few-shot and data-rich regimes, providing a unified foundation for wireless learning.

\end{abstract}


\begin{figure*}[!t]
\centering
\includegraphics[width=0.9\textwidth]{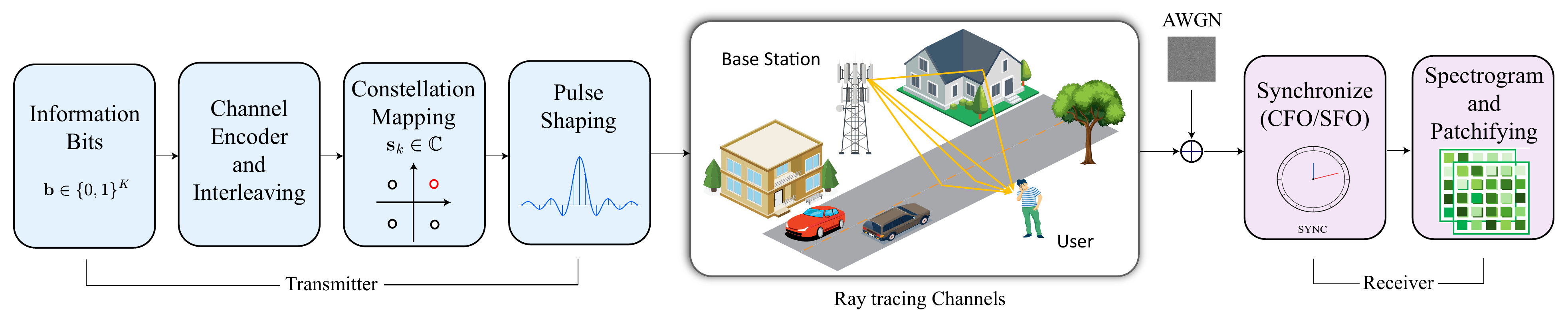}
\caption{Block diagram of the baseband-equivalent spectrogram generation pipeline.
Information bits are encoded, interleaved, mapped, and pulse-shaped to form the transmitted signal $x[n]$, which propagates through a ray-tracing–based fading channel with AWGN.
At the receiver, carrier- and sampling-frequency offsets (CFO/SFO) are compensated, and the resulting baseband signal $y[n]$ is converted to a spectrogram $\mathbf{S}$ via STFT.
Only baseband processing is considered, assuming ideal DAC/ADC.}
\label{fig:block-diagram}
\end{figure*}

\section{Introduction}

The transition toward 6G positions artificial intelligence (AI) as a core enabler of physical-layer processing and autonomous operation, requiring models that generalize across diverse propagation conditions, communication settings, and task domains \cite{survey2021, oshea2017, Morais2024c}. 
While deep learning has achieved strong performance in tasks such as modulation classification and channel estimation \cite{oshea2017, survey2021}, its dependence on large task-specific labeled datasets limits scalability. 
Foundation models address this challenge by leveraging large-scale unlabeled data to learn transferable representations, and Transformer architectures with masked prediction objectives—successful in NLP and vision \cite{devlin2019, dosovitskiy2021}—enable efficient adaptation to downstream tasks. 
In this work, we apply this paradigm to wireless inference using spectrograms of received baseband signals, demonstrating effectiveness on modulation classification and joint SNR/mobility recognition.

The received in-phase and quadrature (I/Q) baseband waveform inherently encodes the physical-layer state of a wireless link, capturing information about the transmitted signal, propagation conditions, and diverse environments~\cite{shun2023tcomm}. Representing these signals as spectrograms via the Short-Time Fourier Transform (STFT) provides a structured time–frequency view that reveals both modulation signatures and channel-induced effects~\cite{shun2023tcomm,wu2022iccc}. Existing spectrogram-based wireless learning methods~\cite{wu2022iccc,shun2023tcomm} rely on supervised training with limited datasets, restricting generalization to new environments, communication standards, or unseen signal conditions and requiring costly labeled data collection for each task. In parallel, recent foundation models for wireless systems~\cite{alikhani2024lwm,aboulfotouh2024} focus primarily on propagation channels or struggle to generalize across heterogeneous communication standards and diverse propagation conditions, such as varying SNR, mobility, and site-specific multipath.

To address these limitations, we introduce \textit{LWM-Spectro}, a transformer-based foundation model pretrained on 9.2 million I/Q spectrograms covering a large number of site-specific propagation environments, along with multiple mobility and SNR conditions.
The model is trained across multiple wireless standards (WiFi, LTE, and 5G) to learn generalizable representations under diverse channel conditions.
To support this diversity within a single model, LWM-Spectro employs a mixture-of-experts (MoE) architecture with protocol-specialized Transformer encoders and a lightweight router that dynamically selects the most relevant expert for each input.
This design enables protocol-aware feature extraction while sharing propagation modeling across experts, reducing active parameters and yielding transferable representations for a broad range of physical-layer tasks. The main contributions of this work are summarized as follows:
\begin{itemize}
\item \textbf{Systematic spectrogram generation for foundation pretraining:} 
We develop a reproducible pipeline for synthesizing diverse I/Q spectrograms across WiFi, LTE, and 5G using protocol-compliant PHY chains, 3GPP channel and mobility models, and SNR sweeps, enabling large-scale pretraining that captures both protocol-specific and propagation-induced structure.

\item \textbf{Unified MoE-based foundation model for wireless spectrograms:}
We pretrain protocol-specific Transformer encoders using masked spectrogram modeling and contrastive learning to obtain transferable time--frequency representations. 
These encoders are then integrated as experts within an MoE framework, where a lightweight router activates the most relevant expert per input, reducing active parameters and achieving substantial few-shot gains over state-of-the-art deep learning models.
\end{itemize}

Together, these contributions establish LWM-Spectro as a scalable foundation model for wireless spectrograms, enabling robust generalization across protocols, propagation environments, and physical-layer tasks.



\section{LWM-Spectro: System Model and Dataset Generation}
\label{sec:system-model}

The objective of LWM-Spectro is to learn transferable spectrogram representations applicable to a wide range of physical-layer tasks. 
Accordingly, the model is pretrained as a foundation model on large-scale data that capture common spectro-temporal structure while generalizing across diverse signal conditions.

This section presents the discrete-time baseband model and the unified dataset generation pipeline used to synthesize large-scale I/Q spectrograms, including protocol-compliant transmit processing, site-specific channel modeling, and time--frequency transformation, with illustrative examples highlighting propagation- and mobility-induced effects.

\subsection{Discrete-Time Baseband System Model}

We consider a single-user, single-antenna, discrete-time baseband communication system operating over a multipath fading channel. A frame of $B$ information bits is denoted by
\begin{equation}
    \mathbf{b} = [b_0,\ldots,b_{B-1}]^\top,\quad b_i \in \{0,1\}.
\end{equation}
The bits are encoded using a channel code with rate 
$R_c = B / N_c$. 
The resulting coded sequence is
\begin{equation}
    \mathbf{c} = f_{\mathrm{enc}}(\mathbf{b})
              = [c_0,\ldots,c_{N_c-1}]^\top .
\end{equation}

The coded bits are mapped to complex modulation symbols drawn from an $M$-ary constellation $\mathcal{S}_M$. With $m=\log_2 M$, each symbol is obtained as
\begin{equation}
    s[i] = f_{\mathrm{mod}}(c_{im},\ldots,c_{(i+1)m-1}),
    \qquad s[i] \in \mathcal{S}_M.
\end{equation}
The discrete-time transmit waveform is generated through pulse shaping using a filter $g[n]$ of length $N_g$ and an oversampling factor $N_{\mathrm{os}}$, i.e., 
\begin{equation}
    x[n] = \sum_{i=0}^{N_s-1} s[i]\, g[n - iN_{\mathrm{os}}], 
    \quad n = 0,\ldots,N_x-1 .
\end{equation}

The received baseband signal follows the time-varying tapped-delay-line (TDL) model
\begin{equation}
    y[n] = \sum_{l=0}^{L-1} h_l[n]\, x[n-l] + w[n],
    \label{eq:tdl_model}
\end{equation}
where $h_l[n]$ is the complex channel coefficient of tap $l$, and $w[n]\sim\mathcal{CN}(0,\sigma_w^2)$ denotes additive white Gaussian noise.

Following the 3GPP TR~38.901 TDL model~\cite{3GPP-TR38.901}, each channel tap is expressed as
\begin{equation}
    h_l[n] = \sqrt{p_l}\,\alpha_l[n],
\end{equation}
where $p_l$ denotes the normalized path power and $\alpha_l[n]$ is a temporally correlated fading process generated according to the 3GPP Doppler spectrum. 
The path delays and powers $\{(\tau_l, p_l)\}$ are obtained from ray-tracing–derived multipath components in the DeepMIMO dataset~\cite{deepmimo2019}. 
Continuous delays $\tau_l$ are discretized and mapped to TDL tap indices, and the corresponding fading sequences are synthesized, resulting in a channel model that jointly captures geometric multipath structure and realistic temporal fading dynamics.

\subsection{Spectrogram Generation via STFT}

The received complex baseband sequence $y[n] = y_I[n] + j\, y_Q[n]$ contains the full in-phase and quadrature information of the propagation channel over time. To construct a joint time--frequency representation, we apply the STFT using an analysis window $w[m]$ of length $N_w$ and hop size $R$ samples. For the $t$-th time frame, the STFT is given by
\begin{equation}
    Y[t,k] = 
    \sum_{m=0}^{N_w-1} y[tR + m]\, w[m]\,
    e^{-j 2\pi k m / N_w},
    \label{eq:stft}
\end{equation}
where $k = 0,\ldots, N_w - 1$ denotes the discrete frequency index and
\begin{equation}
    T = \Big\lfloor \frac{N_x - N_w}{R} \Big\rfloor + 1
\end{equation}
is the total number of frames.

The spectrogram is obtained by taking the squared magnitude of the STFT coefficients:
\begin{equation}
    P[t,k] = |Y[t,k]|^2.
\end{equation}
This results in a two-dimensional time--frequency matrix capturing the instantaneous power distribution of the received signal. Prior to being used as input to a neural network, the spectrogram is log-scaled, normalized, and arranged as
\begin{equation}
    \mathbf{S} = \mathcal{G}(P[t,k]) \in \mathbb{R}^{T \times K \times C},
\end{equation}
where $\mathcal{G}(\cdot)$ denotes the preprocessing pipeline, $K$ is the number of retained frequency bins, and $C$ is the number of channels used in the final representation.

\subsection{An Example and Spectrogram Generation Principle}

\begin{figure}[htbp]
\centering
\includegraphics[width=\columnwidth]{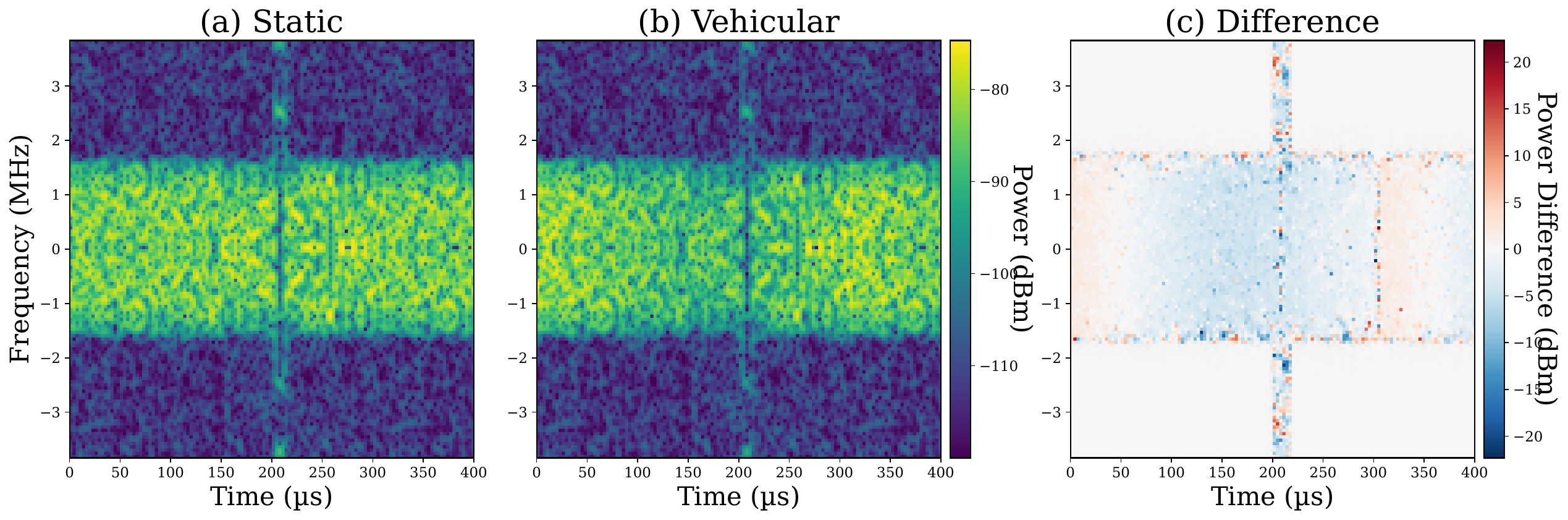}
\caption{WiFi BPSK spectrograms (128 × 128 pixels, coding rate = 1/2, SNR = 20 dB) under static and vehicular mobility.}
\label{fig:mobility-comparison}
\end{figure}

To illustrate the effect of channel dynamics on spectrograms, we generate I/Q signals using NeoRadium~\cite{neoradium-tdl} with 3GPP-compliant site-specific TDL channels parameterized by ray-tracing–derived delays and powers from DeepMIMO. 
Single-channel ($C=1$) power spectrograms are considered. 
Figure~\ref{fig:mobility-comparison} compares WiFi BPSK spectrograms (rate~1/2, SNR~20~dB) under static (0~km/h) and vehicular (30~km/h) mobility, showing Doppler-induced temporal fluctuations and spectral broadening that confirm spectrograms encode propagation dynamics.

With this setup, we generate a large-scale pretraining dataset by simulating received I/Q signals with channel data from DeepMIMO \cite{deepmimo2019}, spanning multiple wireless standards (WiFi, LTE, 5G), modulation and coding schemes (MCS), SNR levels, and mobility regimes. For each city scenario, 1,000 unique channel realizations are sampled, resulting in approximately 460,000 spectrograms per city.
In total, spectrograms from 20 different city scenarios are generated, yielding 9.2 million samples used to pretrain LWM-Spectro across the three communication standards.

\section{Model Architecture and Training Objective}
\label{sec:pretraining}

In this section, we first provide a detailed formulation of the pretraining methodology for LWM-Spectro. 
The model processes spectrogram inputs as sequences of patch embeddings using a Transformer encoder, and is trained with a joint masked spectrogram modeling and contrastive objective to learn robust, discriminative representations. We then discuss the classification head for downstream tasks, along with its training objective and gradient flow.

\subsection{Patch Embedding and Tokenization}

Given a real-valued spectrogram
$\mathbf{S} \in \mathbb{R}^{T \times K}$,
we divide it into non-overlapping $P \times P$ patches \cite{dosovitskiy2021, alikhani2024lwm}. 
The total number of patches is
$N = \frac{T}{P} \times \frac{K}{P}$.
Each patch $i$ is flattened into a vector 
\(\mathbf{s}_i \in \mathbb{R}^{P^2}\) and linearly projected into a \(d\)-dimensional embedding,
\begin{equation}
    \mathbf{x}_i = \mathbf{W}_{\mathrm{emb}} \, \mathbf{s}_i + \mathbf{b}_{\mathrm{emb}}, 
    \qquad i = 1, \ldots, N,
    \label{eq:patch_embed}
\end{equation}
where \( \mathbf{W}_{\mathrm{emb}} \in \mathbb{R}^{d \times P^2} \) and 
\(\mathbf{b}_{\mathrm{emb}} \in \mathbb{R}^d \) are learnable parameters.
The resulting token sequence is
\(\mathbf{X} = [\mathbf{x}_1, \dots, \mathbf{x}_N]^\top \in \mathbb{R}^{N \times d}\).

\subsection{Positional Encoding}

Since patch embeddings are permutation-invariant, we incorporate spatial information using learnable positional encodings. 
Each patch embedding is augmented as
\begin{equation}
    \mathbf{z}_i^{(0)} = \mathbf{x}_i + \mathbf{p}_i,
    \label{eq:pos_encoding}
\end{equation}
where \(\mathbf{p}_i \in \mathbb{R}^d\) is a learnable positional embedding associated with the $i$-th spectrogram patch.
The resulting sequence \(\mathbf{Z}^{(0)} \in \mathbb{R}^{N \times d}\) is used as input to the Transformer encoder.

\subsection{Transformer Encoder}

The encoder consists of \(L\) stacked Transformer blocks, each composed of multi-head self-attention (MSA) and a feed-forward network (FFN), with residual connections and layer normalization.
For the \(l\)-th layer, queries, keys, and values for head \(h\) are computed as
\[
\mathbf{Q}_h = \mathbf{Z}^{(l)} \mathbf{W}_h^Q,\quad
\mathbf{K}_h = \mathbf{Z}^{(l)} \mathbf{W}_h^K,\quad
\mathbf{V}_h = \mathbf{Z}^{(l)} \mathbf{W}_h^V,
\]
where \(\mathbf{W}_h^Q, \mathbf{W}_h^K, \mathbf{W}_h^V \in \mathbb{R}^{d \times d_h}\) and \(d_h = d/H\).
The attention output is
\begin{equation}
    \mathrm{Attn}_h(\mathbf{Z}^{(l)}) 
    = \mathrm{softmax}\!\left(\frac{\mathbf{Q}_h \mathbf{K}_h^\top}{\sqrt{d_h}}\right)\mathbf{V}_h.
\end{equation}
Multi-head outputs are concatenated and projected,
\[
\mathrm{MSA}(\mathbf{Z}^{(l)}) = 
\Big(\bigoplus_{h=1}^H \mathrm{Attn}_h(\mathbf{Z}^{(l)})\Big)\mathbf{W}^O.
\]
With residual connections, the layer update is
\begin{equation}
\mathbf{Z}^{(l+1)} = 
\mathrm{LayerNorm}\big(\tilde{\mathbf{Z}}^{(l)} + \mathrm{FFN}(\tilde{\mathbf{Z}}^{(l)})\big),
\end{equation}
where
\(\tilde{\mathbf{Z}}^{(l)} = \mathrm{LayerNorm}(\mathbf{Z}^{(l)} + \mathrm{MSA}(\mathbf{Z}^{(l)}))\),
and
\(\mathrm{FFN}(\mathbf{u}) = \sigma(\mathbf{u}\mathbf{W}_1 + \mathbf{b}_1)\mathbf{W}_2 + \mathbf{b}_2\),
with GELU activation \(\sigma(\cdot)\).

\subsection{Masked Spectrogram Modeling}

To encourage the encoder to capture global time--frequency structure rather than memorizing local patterns, we randomly mask a subset of patches.
For each spectrogram, a mask set \(\mathcal{M}\subset\{1,\dots,N\}\) is sampled.
The encoder input tokens are defined as
\begin{equation}
\tilde{\mathbf{x}}_i = 
\begin{cases}
\mathbf{x}_i, & i \notin \mathcal{M}, \\
\mathbf{x}_{\texttt{[MASK]}}, & i \in \mathcal{M},
\end{cases}
\end{equation}
where \(\mathbf{x}_{\texttt{[MASK]}}\) is a shared learnable mask embedding.
This masking strategy forces the encoder to infer missing spectro-temporal content from surrounding unmasked patches, thereby promoting contextual and physically meaningful representations of the spectrogram.
The encoder \(f_{\boldsymbol{\theta}}\) maps \(\{\tilde{\mathbf{x}}_i\}\) to latent representations \(\{\mathbf{z}_i\}\).
A lightweight decoder \(g_\phi\) then reconstructs the masked embeddings,
\begin{equation}
\hat{\mathbf{x}}_i = g_\phi(\mathbf{z}_i), \quad i \in \mathcal{M}.
\end{equation}
The reconstruction objective is
\begin{equation}
\mathcal{L}_{\mathrm{recon}} 
= \frac{1}{|\mathcal{M}|}\sum_{i\in\mathcal{M}}
\|\hat{\mathbf{x}}_i - \mathbf{x}_i\|_2^2.
\end{equation}

\subsection{Contrastive Learning}

While masked reconstruction captures spectrogram structure, it does not explicitly enforce discriminative organization across samples. 
We therefore incorporate supervised contrastive learning to shape the embedding space using label information. 
Given an encoder output \(\mathbf{h}_i\), a normalized projection is computed as
\begin{equation}
    \mathbf{z}_i =
    \frac{\mathbf{h}_i \mathbf{W}_p + \mathbf{b}_p}
    {\|\mathbf{h}_i \mathbf{W}_p + \mathbf{b}_p\|_2},
\end{equation}
where \(\mathbf{W}_p \in \mathbb{R}^{d \times d_p}\).
The contrastive loss is defined as
\begin{equation}
\mathcal{L}_{\mathrm{cont}} 
= \sum_{i \in I} \frac{-1}{|P(i)|} 
\sum_{p \in P(i)} 
\log \frac{\exp(\mathbf{z}_i^\top \mathbf{z}_p / \tau)}
{\sum_{a \in A(i)} \exp(\mathbf{z}_i^\top \mathbf{z}_a / \tau)},
\end{equation}
where \(P(i)\) and \(A(i)\) denote positive and remaining samples, respectively.
This objective promotes compact intra-class clustering and inter-class separation, yielding discriminative representations that benefit low-SNR and few-shot learning.

\subsection{Classification Head}

For downstream tasks, we employ a lightweight residual 1D CNN classifier to further model local dependencies in the encoder output.
Given the final encoder representations \(\mathbf{Z}^{(L)} = [\mathbf{z}_1^{(L)}, \ldots, \mathbf{z}_N^{(L)}] \in \mathbb{R}^{N \times d}\), we first apply mean pooling across tokens to obtain a fixed-length sequence representation,
\[
\bar{\mathbf{z}} = \frac{1}{N}\sum_{i=1}^N \mathbf{z}_i^{(L)} .
\]
This representation is then processed by stacked residual 1D convolutional layers, followed by global average pooling (GAP) and a linear classifier to produce the class logits,
\begin{equation}
\hat{\mathbf{y}} =
\mathbf{W}_{\mathrm{cls}} \,
\mathrm{GAP}(\mathbf{h}^{(L_{\mathrm{CNN}})}) +
\mathbf{b}_{\mathrm{cls}},
\end{equation}
where \(\mathbf{h}^{(L_{\mathrm{CNN}})}\) denotes the output of the final CNN block.

\subsection{Training Objective and Fine-tuning Strategy}

Training proceeds in two stages.
During pretraining, the encoder is optimized using only the reconstruction loss,
\begin{equation}
\mathcal{L}_{\mathrm{pretrain}} = \mathcal{L}_{\mathrm{recon}},
\end{equation}
enabling unsupervised learning of general spectrogram structure.
During fine-tuning, we jointly optimize reconstruction and contrastive objectives,
\begin{equation}
\mathcal{L}_{\mathrm{finetune}}
= \lambda_{\mathrm{recon}}\mathcal{L}_{\mathrm{recon}}+
\lambda_{\mathrm{cont}}\mathcal{L}_{\mathrm{cont}}.
\end{equation}
Under this joint objective, the encoder receives complementary reconstructive and discriminative supervision from the reconstruction and contrastive losses, respectively.
The resulting gradient update for the encoder parameter vector $\boldsymbol{\theta}$ is
\begin{equation}
\nabla_{\boldsymbol{\theta}} \mathcal{L}_{\mathrm{total}}
= \lambda_{\mathrm{recon}} \nabla_{\boldsymbol{\theta}} \mathcal{L}_{\mathrm{recon}}
+ \lambda_{\mathrm{cont}} \nabla_{\boldsymbol{\theta}} \mathcal{L}_{\mathrm{cont}} .
\end{equation}

\begin{figure}[t]
\centering
\includegraphics[width=\columnwidth]{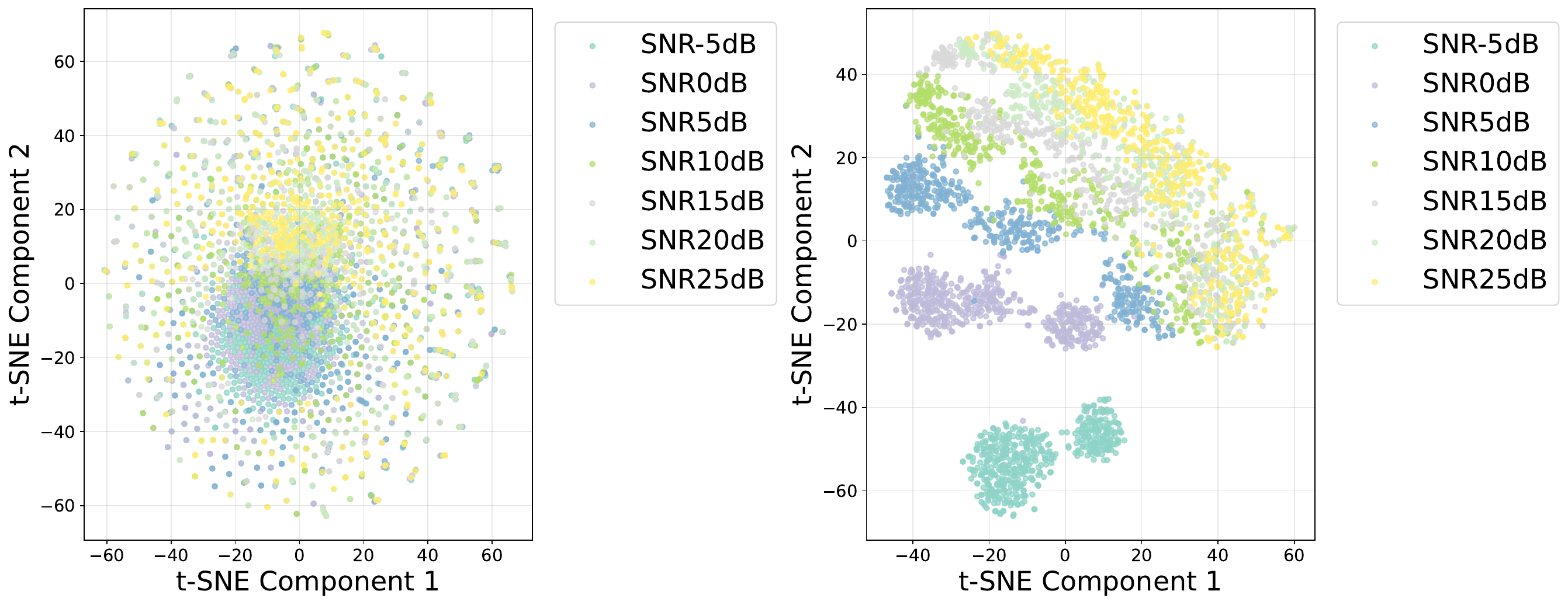}
\caption{t-SNE visualization of 5G samples color-coded by SNR.
Raw spectrograms show heavy overlap, while LWM-Spectro embeddings form well-separated, discriminative clusters by SNR order.}
\label{fig:tsne-comparison}
\end{figure}

\begin{figure}[t]
\centering
\includegraphics[width=\columnwidth]{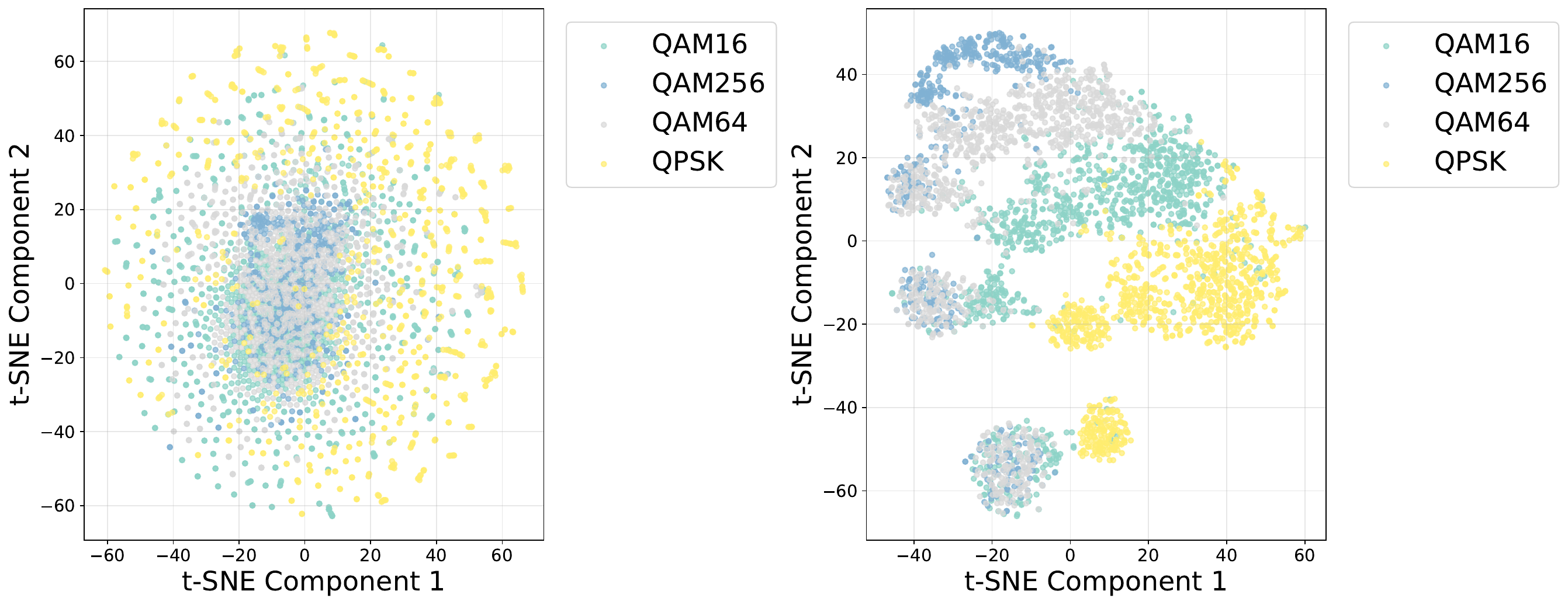}
\caption{t-SNE plot of the same samples color-coded by modulation type.}
\label{fig:tsne-snr-lwm}
\end{figure}

As a result, the encoder learns representations that are simultaneously reconstructive and discriminative.
For downstream tasks such as modulation or Doppler classification, the pretrained encoder \(f_\theta\) is either frozen or fine-tuned with a small learning rate, while task-specific heads are trained on labeled data. Finally, to assess representation quality, we apply t-SNE~\cite{tsne2008} to project learned features into two dimensions. 
Unlike raw spectrograms, which exhibit substantial overlap, LWM-Spectro embeddings organize samples coherently along SNR levels and achieve clear modulation separation at moderate-to-high SNRs (Figures~\ref{fig:tsne-comparison} and~\ref{fig:tsne-snr-lwm}). 
This structure shows that LWM-Spectro learns structured and transferable representations well suited for downstream and few-shot tasks.

\begin{table}[t]
\caption{Pretraining configuration for LWM-Spectro.}
\label{tab:pretraining-config}
\centering
\footnotesize
\resizebox{0.95\linewidth}{!}{
\begin{tabular}{cc}
\hline\hline
\textbf{Component} & \textbf{Setting} \\
\hline
Input representation & $4\times4$ patches; max sequence length 1024 \\
Mask ratio & 70\% \\
Backbone & Transformer (12 layers, $d=128$, 8 heads) \\
Loss weights & $\lambda_{\mathrm{recon}}=1.0$, $\lambda_{\mathrm{cont}}=0.3$ \\
Contrastive temperature & $\tau=0.2$ \\
Optimizer & AdamW ($\beta_1=0.9$, $\beta_2=0.999$, weight decay 0.05) \\
Learning rate schedule & $5\times10^{-4}$, 5-epoch warmup, cosine decay to $10^{-8}$ \\
Training & up to 100 epochs with early stopping \\
\hline\hline
\end{tabular}
}
\end{table}
\section{MoE Architecture}
\label{sec:moe}

This section describes the MoE architecture, detailing its expert encoders, gating network, and aggregation mechanism.

\subsection{Motivation and Design Principles}

Practical systems such as spectrum monitoring and cognitive radio must handle signals from diverse protocols without prior transmitter knowledge \cite{survey2021}. 
While protocol-specific encoders capture standard-dependent time--frequency structure, deploying separate models is inefficient and unscalable, whereas a single shared encoder often entangles heterogeneous features and degrades discrimination. 
The MoE architecture resolves this trade-off by combining protocol-specialized experts $E_k$ with a lightweight router $R$ that selects the most appropriate expert per input, enabling protocol-aware representation learning within a unified and efficient model.

\subsection{Architecture Components}

\paragraph{Expert Encoders}
We employ three protocol-specific encoders $\{E_{\text{WiFi}}, E_{\text{LTE}}, E_{\text{5G}}\}$, each pretrained on its corresponding dataset (Section~\ref{sec:pretraining}). 
Each expert is a 12-layer Transformer with protocol-specific parameters. 
Given an input spectrogram $\mathbf{S}$, the $k$-th expert produces
\[
\mathbf{h}_k = E_k(\mathbf{S}) \in \mathbb{R}^d,\qquad 
k \in \{\text{WiFi}, \text{LTE}, \text{5G}\}.
\]

\paragraph{Gating Network}
The router $R$ maps the input spectrogram to a set of expert-selection weights. 
It is implemented as a 2-layer Transformer encoder ($d = 64$), followed by global average pooling (GAP) and a linear classifier. 
For an encoder output $H \in \mathbb{R}^{T \times K \times d}$, GAP aggregates the time--frequency features into a single $d$-dimensional vector:
\[
\mathrm{GAP}(H) = \frac{1}{TK} \sum_{t=1}^{T} \sum_{k=1}^{K} H_{t,k}.
\]
The router then projects this pooled representation into a three-dimensional logit vector and applies a softmax to obtain the expert weights:
\[
\mathbf{g}
= \mathrm{softmax}\!\left( W_R\,\mathrm{GAP}(R(\mathbf{S})) + \mathbf{b}_R \right)
\in \mathbb{R}^3.
\]
Here, $\mathbf{g} = [g_{\text{WiFi}},\, g_{\text{LTE}},\, g_{\text{5G}}]^\top$ assigns a probability to each protocol-specific expert, determining how strongly each expert contributes to the final mixture.

\paragraph{Expert Aggregation}
We express the MoE output in its general weighted form:
\[
\mathbf{h}_{\text{MoE}} = \sum_{k \in \{\text{WiFi}, \text{LTE}, \text{5G}\}} g_k\, \mathbf{h}_k,
\]
where $g_k$ denotes router-assigned mixing coefficients.  
For efficient deployment, we use a top-1 approximation:
\[
k^* = \arg\max_k g_k, \qquad 
\mathbf{h}_{\text{MoE}} \approx \mathbf{h}_{k^*}.
\]
Thus, the model is formulated as a mixture but evaluated using only the highest-scoring expert.

\subsection{Training Strategy}
The MoE is trained in two stages to decouple expert specialization from routing.
\paragraph{Stage 1: Expert Pretraining}
Each expert encoder $E_k$ is pretrained on its protocol-specific dataset following Section~\ref{sec:pretraining}, enabling it to learn domain-specific features.
\paragraph{Stage 2: Router Training}
With expert weights frozen, only the router $R$ is trained on a balanced mix of WiFi, LTE, and 5G spectrograms to assign mixing weights—yielding a top-1 expert selection at inference.

\subsection{Computational Efficiency}

LWM-Spectro employs top-1 routing, activating only a single expert per input at inference time.
This results in sub-linear scaling with respect to the number of experts:
\begin{itemize}
\item \textbf{Latency:} $\mathcal{O}(C_R + C_E)$, compared to $\mathcal{O}(K \cdot C_E)$ for evaluating all $K$ experts, where $C_R \ll C_E$.
\item \textbf{Memory:} Expert parameters are shared across tasks, avoiding replication of standalone models.
\item \textbf{Throughput:} Inference FLOPs remain nearly constant regardless of protocol distribution.
\end{itemize}
This sparse MoE design enables efficient inference while preserving model capacity.

\section{Applications: Downstream Tasks}
\label{sec:downstream}

We evaluate LWM-Spectro on two downstream tasks using spectrograms from the unseen scenario, ensuring that performance reflects generalization to new pr environments.

\subsection{Task 1: Modulation Classification}

\begin{figure}[t]
\centering
\includegraphics[width=0.9\columnwidth]{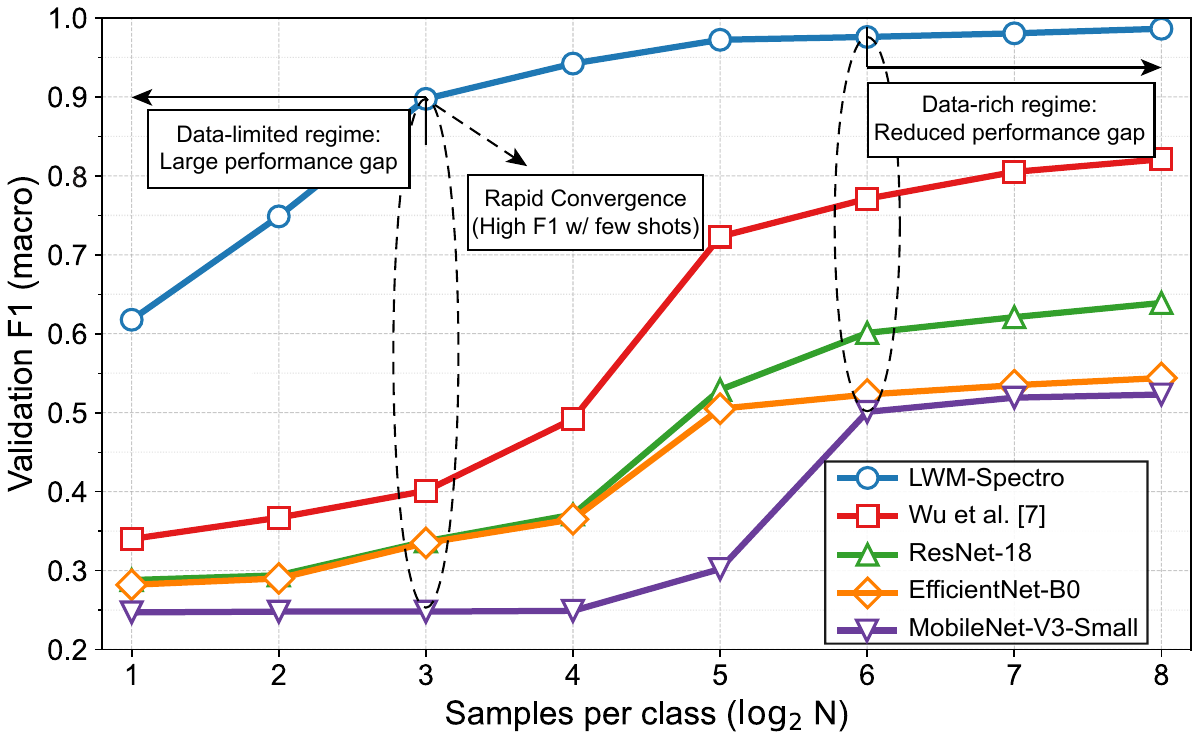}
\caption{Validation macro-F1 score for modulation classification across training set sizes. The LWM embedding sustains higher accuracy than baseline models, particularly in few-shot regimes.}
\label{fig:mcs-f1}
\end{figure}

We assess modulation recognition over five schemes (BPSK, QPSK, 16/64/256-QAM) using spectrograms from the Phoenix scenario in \cite{deepmimo2019}, which is not used for training. For each trial, we sample disjoint tuples, normalize with pretrained statistics, and form stratified train/val/test splits (2--256 samples per class for the downstream task training). Baselines include a re-implemented CNN from Wu et al.~\cite{wu2022iccc} and fine-tuned ImageNet backbones (ResNet-18~\cite{he2016deep}, EfficientNet-B0~\cite{tan2019efficientnet}, MobileNet-V3~\cite{howard2019searching}), all trained with identical optimization settings. As shown in Fig.~\ref{fig:mcs-f1}, LWM maintains a substantial gain in few-shot regimes with only 2–16 samples per class, where CNN-based models perform near chance level. Even with larger training sets, raw-spectrogram baselines only partially close the gap. These results demonstrate that rich pretraining yields transferable features that significantly improve downstream accuracy, particularly in few-shot settings.

\subsection{Task 2: Joint SNR and Doppler Classification}

We evaluate joint SNR and Doppler classification using spectrograms, pairing each SNR bin with two mobility regimes (pedestrian, vehicular) so that every (SNR, mobility) combination becomes a class. We evaluate two strategies: (1) \textbf{LWM\_FT}, jointly fine-tuning encoder and head with reduced learning rates, and (2) \textbf{LWM}, freezing the encoder. Baselines include Deep CNN~\cite{shun2023tcomm} and ImageNet models (\cite{he2016deep}, ~\cite{tan2019efficientnet}, \cite{howard2019searching}). All models use AdamW (weight decay $5\times10^{-4}$) for 8 epochs with early stopping. Table~\ref{tab:task2-results} presents results for LTE protocol. As shown in Table~\ref{tab:task2-results}, LWM\_FT achieves the highest F1 across all data regimes, particularly in few-shot settings where baselines struggle. The frozen-encoder variant (LWM) still outperforms all CNN and ImageNet models, indicating the strength of the pretrained representations. Overall, these results demonstrate that LWM-Spectro effectively captures joint SNR and Doppler characteristics even with limited labeled data.

\begin{table}[t]
\caption{Task 2 validation F1 scores (\%) for joint SNR/Doppler classification on LTE scenario.}
\label{tab:task2-results}
\centering
\footnotesize
\setlength{\tabcolsep}{3pt}
\begin{tabular}{c|cccccc}
\hline\hline
\textbf{N/cls} & \textbf{LWM\_FT} & \textbf{LWM} & \textbf{Deep CNN} & \textbf{R-18} & \textbf{Eff-B0} & \textbf{Mob-V3} \\
\hline
5   & \textbf{76.53} & 47.41 & 44.64 & 28.27 & 30.74 & 7.12 \\
10  & \textbf{91.24} & 52.15 & 46.28 & 30.35 & 35.82 & 10.94 \\
20  & \textbf{92.93} & 78.84 & 50.91 & 40.02 & 38.95 & 27.71 \\
50  & \textbf{93.92} & 89.26 & 53.45 & 41.54 & 39.43 & 30.37 \\
100 & \textbf{94.43} & 90.82 & 82.27 & 48.86 & 40.13 & 36.42 \\
200 & \textbf{94.85} & 91.12 & 90.03 & 49.92 & 41.08 & 37.75 \\
400 & \textbf{95.14} & 92.01 & 90.54 & 50.13 & 42.28 & 39.94 \\
\hline\hline
\end{tabular}
\vspace{-2mm}
\end{table}

\subsubsection{Unified Multi-Protocol Performance}

Table~\ref{tab:moe-results} shows the unified performance when WiFi, LTE, and 5G signals are mixed, requiring dynamic expert selection. The MoE-Router maintains stable 89--90\% accuracy across all training sizes, demonstrating robust multi-protocol generalization.

\begin{table}[t]
\caption{Multi-protocol evaluation on Phoenix scenario with mixed WiFi/LTE/5G signals.}
\label{tab:moe-results}
\centering
\footnotesize
\setlength{\tabcolsep}{5pt}
\begin{tabular}{c|ccc}
\hline\hline
\textbf{Model} & \textbf{N/cls} & \textbf{Accuracy (\%)} & \textbf{F1 (\%)} \\
\hline
\multirow{6}{*}{MoE-Router (3 experts)} 
  & 100   & \textbf{89.79 $\pm$ 1.34} & \textbf{89.77 $\pm$ 1.42} \\
  & 200   & \textbf{90.39 $\pm$ 1.29} & \textbf{90.36 $\pm$ 1.21} \\
  & 400   & \textbf{90.51 $\pm$ 1.18} & \textbf{90.45 $\pm$ 1.27} \\
  & 800   & \textbf{90.84 $\pm$ 1.12} & \textbf{90.81 $\pm$ 1.09} \\
  & 1,600 & \textbf{91.63 $\pm$ 0.98} & \textbf{91.12 $\pm$ 1.06} \\
  & 3,200 & \textbf{92.03 $\pm$ 0.93} & \textbf{91.50 $\pm$ 0.88} \\
\hline
\multirow{6}{*}{Deep CNN~\cite{shun2023tcomm}} 
  & 100   & 43.81 $\pm$ 4.63 & 39.41 $\pm$ 4.78 \\
  & 200   & 47.94 $\pm$ 4.51 & 43.53 $\pm$ 4.37 \\
  & 400   & 56.65 $\pm$ 4.19 & 55.49 $\pm$ 4.06 \\
  & 800   & 64.81 $\pm$ 3.72 & 63.95 $\pm$ 3.89 \\
  & 1,600 & 79.70 $\pm$ 3.11 & 79.61 $\pm$ 3.24 \\
  & 3,200 & 88.64 $\pm$ 2.87 & 88.60 $\pm$ 3.01 \\
\hline
\multirow{6}{*}{ImageNet (R-18~\cite{he2016deep})} 
  & 100   & 44.10 $\pm$ 7.42 & 32.92 $\pm$ 7.18 \\
  & 200   & 45.33 $\pm$ 6.93 & 35.17 $\pm$ 6.74 \\
  & 400   & 47.75 $\pm$ 6.41 & 34.46 $\pm$ 6.92 \\
  & 800   & 49.34 $\pm$ 5.88 & 39.17 $\pm$ 6.25 \\
  & 1,600 & 88.32 $\pm$ 5.32 & 88.29 $\pm$ 5.41 \\
  & 3,200 & 90.11 $\pm$ 4.91 & 90.37 $\pm$ 5.18 \\
\hline\hline
\end{tabular}
\vspace{-2mm}
\end{table}

The MoE-Router reaches 89.8\% F1 with only 100 samples per class—within 98\% of its saturated performance at 3,200 samples—while maintaining low variance across runs (Table~\ref{tab:moe-results}). In contrast, the Deep CNN~\cite{wu2022iccc} requires over 30$\times$ more data to approach 90\% F1 and exceeds 80\% F1 only beyond 1,600 samples. ImageNet-based models also perform poorly in few-shot regimes, reflecting the domain gap between natural images and wireless spectrograms.

\section{Concluding Remarks}
\label{sec:conclusion}

We introduced LWM-Spectro, a transformer-based foundation model for I/Q spectrograms pretrained on 9.2 million WiFi, LTE, and 5G samples. Using masked spectrogram modeling, contrastive learning, and a mixture-of-experts framework, the model learns robust time–frequency representations that capture both protocol-specific and channel-induced features. Across modulation and joint SNR/mobility classification tasks, LWM-Spectro shows strong transferability and substantial few-shot gains over the state-of-the-art deep learning baselines, demonstrating the effectiveness of large-scale self-supervised pretraining for wireless physical-layer learning.

\vfill

\bibliographystyle{IEEEtran}
\bibliography{references, AlkhateebRefs}

\end{document}